\documentclass{aa} 
\usepackage{txfonts}
\usepackage{graphicx}
\usepackage{mathrsfs}
\usepackage[dvips]{rotating}
\usepackage{natbib}
\bibpunct{(}{)}{;}{a}{}{,}

\def\kms{$\mathrm{km}~\mathrm{s}^{-1}$}
\def\CI{\hbox{C\,{\sc i}}}
\def\NI{\hbox{N\,{\sc i}}}
\def\OI{\hbox{O\,{\sc i}}}

\def\FeI{\hbox{Fe\,{\sc i}}}
\def\FeII{\hbox{Fe\,{\sc ii}}}
\def\MgI{\hbox{Mg\,{\sc i}}}
\def\MgII{\hbox{Mg\,{\sc ii}}}
\def\SiI{\hbox{Si\,{\sc i}}}
\def\SiII{\hbox{Si\,{\sc ii}}}
\def\CaI{\hbox{Ca\,{\sc i}}}
\def\CaII{\hbox{Ca\,{\sc ii}}}
\def\TiI{\hbox{Ti\,{\sc i}}}
\def\TiII{\hbox{Ti\,{\sc ii}}}
\def\SrII{\hbox{Sr\,{\sc ii}}}
\def\NaI{\hbox{Na\,{\sc i}}}
\def\AlI{\hbox{Al\,{\sc i}}}
\def\ScII{\hbox{Sc\,{\sc ii}}}

\def\CrII{\hbox{Cr\,{\sc ii}}}
\def\MnI{\hbox{Mn\,{\sc i}}}

\def\ZnI{\hbox{Zn\,{\sc i}}}
\def\YII{\hbox{Y\,{\sc ii}}}
\def\ZrII{\hbox{Zr\,{\sc ii}}}
\def\BaII{\hbox{Ba\,{\sc ii}}}
\def\HeI{\hbox{He\,{\sc i}}}
\def\MnII{\hbox{Mn\,{\sc ii}}}
\def\AlII{\hbox{Al\,{\sc ii}}}
\def\CuI{\hbox{Cu\,{\sc i}}}
\def\LaII{\hbox{La\,{\sc ii}}}
\def\NdII{\hbox{Nd\,{\sc ii}}}

\begin{document}

\title{Chemical analysis of 24 dusty (pre-)main-sequence stars}
\author{B.~Acke \and C.~Waelkens}
\authorrunning{Acke \& Waelkens}
\titlerunning{Elemental abundances of young stars}
\institute{Instituut voor Sterrenkunde, Katholieke Universiteit Leuven}
\date{DRAFT, \today}

\abstract{{We have analysed the chemical photospheric composition of
    24 Herbig Ae/Be and Vega-type stars in search for the $\lambda$ Bootis
    phenomenon. We present the results of the elemental abundances of the
    sample stars. Some of the stars were never before studied
    spectroscopically at optical wavelengths.  We have determined the
    projected rotational velocities of our sample stars. 
    Furthermore, we discuss stars that depict a (selective) depletion
    pattern in detail. \object{HD 4881} and \object{HD 139614} seem to display an
    overall deficiency. \object{AB Aur} and possibly \object{HD 126367} have
    subsolar values for the iron abundance, but are almost solar in
    silicon. \object{HD 100546} is the only clear $\lambda$ Bootis star in our
    sample. }  
\keywords{stars: abundances -- stars: chemically peculiar -- planetary
    systems: protoplanetary disks -- stars: pre-main-sequence --
    stars: individual: \object{HD 100546}}} 

\maketitle

\section{Introduction}

Herbig Ae/Be (HAEBE) stars are generally thought to be the somewhat more
massive analogues of the T Tauri stars, which are low-mass
pre-main-sequence stars. They were first defined as a class in 1960 by
\citeauthor{herbig60}, based on three observational criteria: the stars have spectral
type A or earlier, display emission lines, and illuminate a bright
reflection nebula in their immediate vicinity. Later on other
observational criteria were proposed. \citet{finkenzeller84}
mentioned that a shared characteristic of Herbig Ae/Be stars is the
presence of an IR excess, which is an independent, but more easily
detected signature of circumstellar matter. Following \citet{the} a
useful working definition is: {\it Herbig Ae/Be stars have 
  spectral type B to F8 and luminosity class III to V, and show
  emission lines. They display a flux excess in the IR, due to the
  presence of cool and/or hot circumstellar matter.} The observational
properties of HAEBE stars have been reviewed by e.g. \citet{perez} and
\citet{waters98}.
 
That HAEBE stars are young objects has been convincingly confirmed by
their presence in star-forming regions and also by the location in the
Hipparcos HR diagram of individual objects
\citep[e.g.][]{vandenancker98}. Some of the objects often classified as HAEBE
stars may be young main-sequence stars rather than pre-main-sequence
stars, which is not really surprising, since, the more massive the
star, the closer the circumstellar dissipation timescale approaches
the pre-main-sequence lifetime. Dusty IR excesses, but without
accompanying gaseous emission lines, are also observed in slightly or
more evolved main-sequence stars, of which $\beta$ Pictoris and Vega
are the most well known examples. The generic link between these
debris-disk stars and HAEBE stars is a subject of current research to
which we aim to contribute with the present paper.  

Vega is the prototype of the Vega-like stars and shows mild
characteristics of the so-called $\lambda$ Bootis stars, which are
chemically peculiar A/F main sequence stars: the elements C, N, O and
S have approximately solar photospheric abundances, while several
metals are depleted. \citet{venn90} pointed out that the
$\lambda$ Boo abundance pattern is comparable to the composition of
the gas component of the interstellar medium \citep[e.g.][]{heiter02}: the
depleted elements (Mg, Ca, Ti, Fe and Sr) are those with the higher
condensation temperatures. They suggested that the peculiar abundances
could arise from gas/dust separation in the circumstellar environment,
followed by accretion of the metal depleted gas. Reaccretion of
circumstellar gas which is `cleaned' by this mechanism explains the
most peculiar chemical composition of several post-AGB stars, for
which the depletion pattern is much more pronounced than in the
$\lambda$ Bootis stars \citep[e.g.][]{lambert88, waters92}. These
peculiar post-AGB stars all turn out to be members of binaries, and
the circumstellar dust occurs in a long-lived disk which surrounds the
binary system \citep[e.g.][]{vanwinckel95, waters98b, molster99}. 

Observational evidence
\citep[e.g.][]{mannings97,testi,pietu,fuente,natta04} shows that the 
circumstellar matter in the vicinity of HAEBE 
stars displays a disk-like geometry, the modelling of which currently
is the subject of intense research (e.g. disks powered by accretion
\citep{lin80,bell94}, irradiated passive disks \citep{kenyon87},
irradiated flared disks \citep{chiang}, disks with an inner hole and a
puffed up inner rim \citep{dullemond01,dullemond02,dullemond04}). From
the observation of redshifted resonance 
lines in the spectra of HAEBE stars, evidence for matter accretion has
been found. These stars then seem natural candidates for displaying
the $\lambda$ Bootis phenomenon. With the aim of further exploring the
possible evolutionary connection between the Herbig Ae/Be stars and
the Vega-type stars, we have carried out a chemical analysis of a
selection of 24 objects, consisting of confirmed HAEBE stars and
debris-disk stars and other objects which may be of interest in this
context.  

The plan of the paper is as follows: in the next section we describe
the sample and the observations; Section~\ref{results} is devoted to the general
results from the chemical analysis and a specific discussion of the
objects for which depletion may occur. In Section~\ref{discussion} we
discuss the results. In the
Appendix, some notes on the individual objects are summarized, as well
as a comparison with literature results.

\section{Observations and Analysis}

\begin{table}
\caption{ The program stars, listed according to their HD
  numbers. The
  effective temperature and gravity of the Kurucz model used in the
  present abundance analysis. The last column indicates the category
  to which the objects belong. References: a--\citet{backman93};
  b--\citet{bhatt00}; c--\citet{coulson98}; d--\citet{kalas02};
  e--\citet{malfait}; f--\citet{patten91}; g--\citet{song00};
  h--\citet{the}. \label{programstars}}
\begin{center}
\begin{tabular}{ccccc}
 \hline
 \multicolumn{5}{c}{\bf Program stars}\\ \hline \hline
 \multicolumn{2}{c}{Object} &
 \multicolumn{1}{c}{$T_{\mathrm{e}}$} &
 \multicolumn{1}{c}{$\log g$} &
 \multicolumn{1}{c}{Cat.} \\  
 \multicolumn{1}{c}{HD} &
 \multicolumn{1}{c}{Other} &
 \multicolumn{1}{c}{[K]} &
 \multicolumn{1}{c}{[log cm/s]} &
 \multicolumn{1}{c}{}  \\ 
\hline 
\object{HD 4881}   & \object{HR 241}        &  11000  & 4.5 & VEGA$^d$  \\                      
\object{HD 6028}   & \object{HR 287}        &   8750  & 3.5 & VEGA$^c$  \\		     
\object{HD 17081}  & \object{$\pi$ Cet}     &  13000  & 4.0 & HAEBE$^e$ \\		     
\object{HD 17206}  & \object{$\tau_{1}$ Eri}&   6500  & 4.0 & VEGA$^b$  \\		
\object{HD 18256}  & \object{$\rho$ Ari}    &   6750  & 4.5 & Check Star\\
\object{HD 20010}  & \object{$\alpha$ For}  &   6500  & 4.5 & VEGA$^b$  \\		     
\object{HD 28978}  & \object{HR 1448}       &   9250  & 4.0 & VEGA$^a$  \\			     
\object{HD 31293}  & \object{AB Aur}        &   9750  & 5.0 & HAEBE$^h$ \\		     
\object{HD 31648}  &                        &   8750  & 4.0 & HAEBE$^h$ \\
\object{HD 33564}  & \object{HR 1686}       &   6250  & 4.0 & VEGA$^f$  \\
\object{HD 36112}  &                        &   7750  & 3.5 & HAEBE$^h$ \\		     
\object{HD 95418}  & \object{$\beta$ UMa}   &   9750  & 4.0 & VEGA $^{b,\, g}$ \\			     
\object{HD 97048}  & \object{CU Cha}        &  10000  & 4.0 & HAEBE$^h$ \\					     
\object{HD 97633}  & \object{$\theta$ Leo}  &   9250  & 3.5 & Check Star\\
\object{HD 100453} &                        &   7500  & 4.0 & HAEBE$^e$ \\		     
\object{HD 100546} & \object{KR Mus}        &  11000  & 4.5 & HAEBE$^h$ \\
\object{HD 102647} & \object{$\beta$ Leo}   &   8500  & 4.0 & VEGA$^g$  \\
\object{HD 104237} & \object{DX Cha}        &   8000  & 4.5 & HAEBE$^h$ \\		     
\object{HD 135344} &                        &   6750  & 4.0 & HAEBE$^h$ \\		     
\object{HD 139614} &                        &   8000  & 4.5 & HAEBE$^e$ \\		     
\object{HD 190073} &                        &   9250  & 3.5 & HAEBE$^h$ \\			     
\object{HD 244604} &                        &   8750  & 4.0 & HAEBE$^h$ \\		     
\object{HD 245185} &                        &   9000  & 4.0 & HAEBE$^h$ \\		     
\object{HD 250550} &                        &  11000  & 4.0 & HAEBE$^h$ \\                                                 
 \hline
\end{tabular}
\end{center}
\end{table}

We have obtained spectra for the 24 objects which are listed in Table
\ref{programstars}. Fourteen program stars are HAEBE stars, eight
objects are confirmed Vega-type stars or occur in lists of Vega-type
candidates, and two stars (\object{HD 18256} and \object{HD 97633}) are
rather normal main-sequence A-type stars which we measured to
check our analysis with respect to literature results. A first set of
observations, concerning mostly the northern-hemisphere objects, was
carried out by drs. Hans Van Winckel and Gwendolyn Meeus during the
nights of 23 and 25 
December 1996, with the William Herschel Telescope and the Utrecht
Echelle Spectrograph at La Palma Observatory (Canary Islands, Spain). The second
observational run, concerning mostly southern-hemisphere objects, was
performed by dr. Gwendolyn Meeus from January 26 till January 28, 1999
at the ESO site La Silla (Chile). These data were collected using
the 1.5m telescope and the FEROS instrument. The wavelength range of
the spectra lies between 3860 and 9110\AA. The spectral resolution of
both instruments is comparable ($\mathscr{R} \sim 45,000$). We reduced the data with
the MIDAS software package.

Despite the broad wavelength coverage, for several program stars only
a few lines turned out to be appropriate for detailed abundance
analyses. Such a circumstance is not unusual for A/B-type stars, where
the condition that spectral lines with equivalent widths higher than
150 m\AA\ should be avoided in abundance analyses and the fairly high
projected rotational velocities severely limit the amount of suitable
lines. For a few slowly rotating stars (e.g. \object{HD 18256},
\object{HD 20010}, \object{HD 104327} and \object{HD 139614}) we were
able to use hundreds of lines in the 
analysis; the average value for the amount of lines is about 100. It
has to be noted that the vast majority of usable lines are iron and
titanium lines; our search for metal depletion primarily relies on
these two elements, also on silicon and calcium for most objects, but
on magnesium only in a minority of cases. On the other hand, for 6
program stars, no carbon, nitrogen and oxygen (CNO) abundances could
be determined; the relative scarcity of CNO lines and the non-LTE
(local thermodynamic equilibrium) sensitivity of especially
nitrogen implies that the uncertainties on the CNO abundances are
sometimes critical for assessing whether selective depletion occurs;
in all cases at most a few lines are available for the determination
of the sulphur abundance.

\begin{table*}[!bht]
\caption{The solar abundances for C, N, O and S are respectively 8.56,
  8.05, 8.93 and 7.21 dex \citep{anders89}. The given error represents
  the line-to-line scatter. The number of lines that were used in the abundance
  determination is mentioned in parentheses. \label{CNOS}} 
\begin{center}
\begin{tabular}{ccccc}
 \hline
 \multicolumn{5}{c}{\bf CNO and S abundances}\\ \hline \hline
 \multicolumn{1}{c}{Object} &
 \multicolumn{1}{c}{[C]} &
 \multicolumn{1}{c}{[N]} &
 \multicolumn{1}{c}{[O]} & 
 \multicolumn{1}{c}{[S]} \\
 \multicolumn{1}{c}{HD} &
 \multicolumn{1}{c}{$dex$} &
 \multicolumn{1}{c}{$dex$} &
 \multicolumn{1}{c}{$dex$} &
 \multicolumn{1}{c}{$dex$} \\
\hline
\object{HD 4881}   &$<-$0.39         (3)  & $-$0.49$\pm$0.05 (4)& $-$0.37$\pm$0.10 (3)&                      \\   
\object{HD 17081}  &                      &                     & $-$0.18$\pm$0.03 (4)&$-$0.07 (1)           \\
\object{HD 17206}  & $-$0.10$\pm$0.08 (8) & $-$0.16$\pm$0.02 (2)& $-$0.11$\pm$0.05 (5)&$+$0.00$\pm$0.14 (4)  \\
\object{HD 18256}  & $-$0.10$\pm$0.08 (7) &                     &                     &$-$0.03$\pm$0.03 (2)  \\
\object{HD 20010}  & $-$0.32$\pm$0.16 (11)&                     &                     &$-$0.35$\pm$0.19 (8)  \\
\object{HD 28978}  & $-$0.20$\pm$0.01 (2) &                     & $+$0.00 (1)         &                      \\    
\object{HD 31293}  &                      & $+$0.14 (1)         & $+$0.21$\pm$0.07 (2)&                      \\    
\object{HD 31648}  & $-$0.23$\pm$0.05 (4) &                     &                     &                      \\
\object{HD 33564}  & $+$0.30$\pm$0.12 (4) &                     &                     &$+$0.10$\pm$0.01 (2)  \\
\object{HD 36112}  & $-$0.14$\pm$0.05 (3) & $-$0.01 (1)         &                     &$+$0.16 (1)           \\
\object{HD 95418}  &                      & $-$0.29$\pm$0.02 (2)& $-$0.33$\pm$0.04 (3)&                      \\  
\object{HD 97048}  &                      & $+$0.28 (1)         & $-$0.11$\pm$0.14 (6)&                      \\ 
\object{HD 97633}  &                      & $-$0.29$\pm$0.06 (7)& $-$0.33$\pm$0.01 (2)&$+$0.23$\pm$0.04 (3)  \\
\object{HD 100453} & $+$0.00$\pm$0.09 (7) & $-$0.17 (1)         &                     &$-$0.11$\pm$0.07 (4)  \\
\object{HD 100546} &                      & $+$0.06 (1)         &$-$0.05$\pm$0.10 (10)&                      \\    
\object{HD 104237} & $-$0.14$\pm$0.20 (30)& $-$0.42$\pm$0.15 (6)&$-$0.10$\pm$0.14 (11)&$-$0.15$\pm$0.14 (5)  \\
\object{HD 135344} & $+$0.16 (1)          &                     &                     &                      \\    
\object{HD 139614} & $-$0.06$\pm$0.14 (21)& $-$0.19$\pm$0.03 (4)& $-$0.24$\pm$0.01 (2)&$-$0.22$\pm$0.04 (4)  \\
\object{HD 190073} & $-$0.07$\pm$0.03 (3) & $+$0.59 (1)         & $-$0.27$\pm$0.07 (4)&                      \\
\hline
\object{Vega}      & $-$0.10              & $-$0.05             & $+$0.08             &                      \\
\object{$\lambda$ Boo} & $-$0.37          & $+$0.06             & $-$0.45             &                      \\
ISM                & $-$0.03              & $-$0.01             & $-$0.02             &                      \\
 \hline
\end{tabular}
\end{center}
\end{table*}

As starting values for the stellar parameters (effective temperature
$T_{\mathrm{e}}$ and gravity $\log g$), we have used previously
published results or values derived from photometric data. The
equivalent widths derived from the spectra were converted to
abundances using the program MOOG \citep{sneden73} for cooler stars
($T_{\mathrm{e}} <$ 8500K) and the Kurucz program WIDTH6 for hotter stars
($T_{\mathrm{e}} >$ 8500K), both programs supposing LTE. Non-LTE
effects have not been taken into account. We have computed abundances 
with both programs for the stars with effective temperatures in the
range from 7500K and 9500K, and found good agreement. The stellar
parameters were refined --when necessary and/or possible-- in the
usual way: the effective temperature was determined by asking that the
abundances be independent from the excitation potential for ions with
many observed lines; the gravity was set so that analyses for
different ionisation stages for iron led to the same result. The
stellar parameters as we adopted them in our abundance analyses are
listed in Table~\ref{programstars}. We were limited in chosing the
effective temperature and gravity parameters by the finite grid of
Kurucz models at our disposal --- the grid has a step of 250K in
$T_{\mathrm{e}}$ and a step of 0.5 dex in $\log g$. In each
model, we have used a turbulent velocity of $v_{tur} = 3$ \kms.

\section{Results of the chemical analysis \label{results}}

\subsection{Overview}

In this section we present the derived elemental abundances. All
values are given in reference to the (adopted) solar abundances given
by \citet{anders89}, $[X]=+0.0$ indicating a solar abundance
for element X.  
Since we are investigating whether the selective depletion pattern is
present in our sample stars, we compare the computed abundances with
the elemental abundances of \object{$\lambda$ Boo} and \object{Vega}, two prototype
$\lambda$ Bootis stars, and the interstellar medium
\citep[][]{paunzen99,venn90,qiu01,erspamer02,jenkins89}. The stellar
parameters for \object{$\lambda$ Boo} are $T_{\mathrm{e}}$=8650K and $\log g$=4.00
\citep{venn90} and for \object{Vega} $T_{\mathrm{e}}$=9430K and $\log g$=3.95 \citep{qiu01}.
In Table~\ref{CNOS} we have summarized the measured carbon, nitrogen,
oxygen and sulphur abundances.
Table~\ref{metals} contains the abundances of the metals Mg, Si, Ca, Ti, Fe and Sr. 
In Table~\ref{others}, we have tabulated the abundances we computed
for other elements. In parenthesis, the number of spectral
lines that were used in the abundance
determination are indicated\footnote{The complete linelist and
  measured equivalent widths can be downloaded at \\  
{\bf http://www.ster.kuleuven.ac.be/$\sim$bram/LINELIST/linelist.ascii}
}. All abundances tabulated in Tables~\ref{CNOS}, \ref{metals}
and \ref{others} and mentioned in the text have a line-to-line scatter
smaller than 0.20 dex. The typical standard deviation is $\Delta\sim
0.10$ dex. The uncertainty on the stellar parameters introduces in general
another $\sim$0.10 dex.

\begin{sidewaystable*}
\caption{The abundances of the 'metals'. The adopted solar abundances
  for Mg, Si, Ca, Ti, Fe and Sr are respectively 7.58, 7.55, 6.36,
  4.99, 7.52 and 2.90 dex \citep{anders89}. The given error represents
  the line-to-line scatter. The number of lines that were used in the abundance
  determination is mentioned in parentheses. \label{metals}}  
\scriptsize
\begin{center}
\begin{tabular}{c|cccccccccccc}
 \hline
 \multicolumn{12}{c}{\bf Metal abundances}\\ \hline \hline
 \multicolumn{1}{c|}{ Object} &
 \multicolumn{1}{|c}{ [\MgI]} &
 \multicolumn{1}{c}{ [\MgII]} &
 \multicolumn{1}{c}{ [\SiI]} &
 \multicolumn{1}{c}{ [\SiII]} &
 \multicolumn{1}{c}{ [\CaI]} &
 \multicolumn{1}{c}{ [\CaII]} &
 \multicolumn{1}{c}{ [\TiI]} &
 \multicolumn{1}{c}{ [\TiII]} &
 \multicolumn{1}{c}{ [\FeI]} &
 \multicolumn{1}{c}{ [\FeII]} &
 \multicolumn{1}{c}{ [\SrII]} \\
 \multicolumn{1}{c|}{ HD} &
 \multicolumn{11}{|c}{ in $dex$} \\
\hline
\object{HD 4881}   &                     & $-$0.70$\pm$0.05 (3)&                      & $-$1.15$\pm$0.07 (5)&                      &             &                     & $-$0.49$\pm$0.14 (12)& $-$0.61$\pm$0.10 (4)  & $-$0.72$\pm$0.18 (19)& $-$0.34$\pm$0.09 (2) \\
\object{HD 6028}   &                     &                     &                      & $-$0.23$\pm$0.13 (2)&                      &             &                     & $+$0.11$\pm$0.09 (3) & $-$0.13$\pm$0.01 (3)  & $-$0.14$\pm$0.08 (7) & \\
\object{HD 17081}  &                     & $-$0.11$\pm$0.02 (4)&                      & $+$0.05$\pm$0.28 (4)&                      &             &                     & $-$0.18$\pm$0.13 (7) &                       & $-$0.26$\pm$0.19 (29)& \\
\object{HD 17206}  & $-$0.15 (1)         &                     & $+$0.23$\pm$0.10 (14)& $-$0.04$\pm$0.04 (3)& $+$0.12$\pm$0.05 (3) &             &                     & $-$0.57 (1)          & $-$0.06$\pm$0.18 (81) & $-$0.10$\pm$0.13 (13)& \\
\object{HD 18256}  & $+$0.07$\pm$0.01 (2)&                     & $+$0.09$\pm$0.11 (18)& $+$0.03$\pm$0.02 (2)& $+$0.19$\pm$0.18 (17)&             & $+$0.22$\pm$0.21 (4)& $+$0.16$\pm$0.10 (4) & $+$0.05$\pm$0.13 (115)& $+$0.07$\pm$0.19 (21)& \\
\object{HD 20010}  & $-$0.20$\pm$0.11 (2)&                     & $-$0.08$\pm$0.10 (22)& $-$0.15$\pm$0.01 (2)& $-$0.10$\pm$0.17 (23)&             & $-$0.02$\pm$0.10 (5)& $-$0.06$\pm$0.02 (3) & $-$0.20$\pm$0.16 (146)& $-$0.12$\pm$0.17 (27)& \\
\object{HD 28978}  & $-$0.35 (1)         & $-$0.34 (1)         &                      & $+$0.05$\pm$0.15 (3)& $-$0.06$\pm$0.05 (3) &             &                     & $+$0.21$\pm$0.18 (43)& $+$0.17$\pm$0.14 (45) & $+$0.20$\pm$0.16 (29)& $+$0.31$\pm$0.01 (2)\\
\object{HD 31293}  &                     &                     &                      & $-$0.11$\pm$0.03 (3)&                      &             &                     &                      & $-$1.04$\pm$0.12 (2)  &                      & \\
\object{HD 31648}  & $-$0.53 (1)         &                     & $+$0.71 (1)          &                     & $+$0.11 (1)          &             &                     &                      & $+$0.12 (1)           & $+$0.01 (1)          & \\
\object{HD 33564}  &                     &                     &                      & $+$0.37$\pm$0.03 (2)& $-$0.13$\pm$0.10 (3) &             &                     & $-$0.10$\pm$0.23 (28)& $-$0.35$\pm$0.18 (30) & $-$0.41$\pm$0.19 (13)& \\
\object{HD 36112}  &                     &                     & $+$0.07$\pm$0.11 (3) &                     & $+$0.05$\pm$0.09 (8) &             &                     & $-$0.23$\pm$0.04 (4) & $-$0.12$\pm$0.13 (32) & $-$0.15$\pm$0.11 (11)& \\
\object{HD 95418}  & $-$0.40 (1)         &                     &                      & $+$0.13$\pm$0.05 (2)&                      &             &                     & $+$0.28 (1)          & $+$0.21$\pm$0.07 (7)  & $+$0.32$\pm$0.16 (15)& \\  
\object{HD 97048}  &                     &                     &                      & $-$0.75$\pm$0.06 (2)&                      &             &                     &                      &                       &                      & \\
\object{HD 97633}  &                     &                     &                      & $+$0.14$\pm$0.12 (2)& $-$0.09$\pm$0.12 (3) &             &                     & $+$0.01 (1)          & $+$0.00$\pm$0.10 (17) & $+$0.04$\pm$0.15 (14)& \\  
\object{HD 100453} &                     & $-$0.01$\pm$0.09 (2)& $+$0.13$\pm$0.07 (5) & $+$0.15$\pm$0.04 (2)& $+$0.03$\pm$0.12 (16)&             &                     & $-$0.14$\pm$0.18 (11)& $-$0.10$\pm$0.19 (81) & $-$0.07$\pm$0.19 (26)& \\  
\object{HD 100546} & $-$1.14$\pm$0.07 (4)&                     &                      & $-$0.84$\pm$0.13 (6)&                      &             &                     &                      &                       & $-$1.30$\pm$0.13 (5) & \\  
\object{HD 102647} &                     &                     &                      &                     & $+$0.12$\pm$0.14 (2) &             &                     &                      & $-$0.27$\pm$0.15 (6)  & $-$0.18$\pm$0.11 (4) & \\  
\object{HD 104237} & $-$0.12 (1)         & $-$0.19$\pm$0.07 (2)& $+$0.02$\pm$0.11 (14)& $+$0.31$\pm$0.09 (5)& $+$0.15$\pm$0.12 (21)& $+$0.14 (1) &                     & $+$0.23$\pm$0.19 (36)& $+$0.08$\pm$0.18 (175)& $+$0.10$\pm$0.17 (56)& \\  
\object{HD 135344} &                     &                     & $+$0.21 (1)          & $-$0.05$\pm$0.10 (2)& $+$0.00$\pm$0.27 (4) &             &                     & $-$0.45$\pm$0.01 (2) & $-$0.07$\pm$0.10 (20) & $-$0.09 (1)          & \\  
\object{HD 139614} & $-$0.48$\pm$0.06 (2)& $-$0.35$\pm$0.10 (3)& $-$0.36$\pm$0.04 (2) & $-$0.21$\pm$0.08 (5)& $-$0.24$\pm$0.10 (17)&             &                     & $-$0.13$\pm$0.16 (38)& $-$0.30$\pm$0.15 (116)& $-$0.24$\pm$0.16 (33)& \\  
\object{HD 190073} & $-$0.06 (1)         & $-$0.06$\pm$0.02 (3)&                      & $+$0.08$\pm$0.12 (3)& $+$0.29$\pm$0.09 (3) &             &                     & $+$0.04$\pm$0.13 (26)& $+$0.07$\pm$0.20 (44) & $-$0.07$\pm$0.11 (16)& \\  
\object{HD 244604} &                     &                     &                      &                     &                      &             &                     &                      & $-$0.25 (1)           &                      & \\  
\object{HD 250550} &                     & $-$0.42 (1)         &                      & $-$0.17$\pm$0.09 (3)&                      &             &                     & $+$0.02 (1)          &                       & $-$0.88$\pm$0.12 (4) & \\ 
\hline        	                                              	             	                               
\object{Vega}          & $-$0.77&$-$0.89& & $-$0.59& $-$0.95& $-$1.34& & $-$0.41& $-$0.58& $-$0.59& $-$1.62 \\ 
\object{$\lambda$ Boo} & $-$2.00&$-$2.00& & $-$1.00&        & $-$1.97& & $-$2.03& $-$2.05& $-$2.05& $-$2.10 \\ 
 ISM                   & $-$0.70&$-$0.70& &        & $-$3.60& $-$3.60& & $-$2.70& $-$2.10& $-$2.10&         \\ 
\hline
\end{tabular}
\end{center}
\normalsize
\end{sidewaystable*}

\begin{sidewaystable*}
\caption{The abundances of other elements. The adopted solar
  abundances for Na, Al, Sc, V, Cr, Mn, Ni, Zn, Y, Zr and Ba are
  respectively 6.33, 6.47, 3.10, 4.00, 5.67, 5.39, 6.25, 4.60, 2.24,
  2.60 and 2.13 dex \citep{anders89}. The given error represents
  the line-to-line scatter. The number of lines that were used in the abundance
  determination is mentioned in parentheses. Two numbers refer to the
  neutral and first ionization state of the element, respectively. \label{others}} 
\scriptsize		                				       
\begin{center}
\begin{tabular}{c|ccccccccccc}
 \hline
 \multicolumn{12}{c}{\bf Other elemental abundances}\\ \hline \hline
 \multicolumn{1}{c|}{ Object} &
 \multicolumn{1}{|c}{ [\NaI]} &
 \multicolumn{1}{c}{ [\AlI]} &
 \multicolumn{1}{c}{ [\ScII]} &  
 \multicolumn{1}{c}{ [V]} &
 \multicolumn{1}{c}{ [Cr]} &
 \multicolumn{1}{c}{ [\MnI]} &
 \multicolumn{1}{c}{ [Ni]} &
 \multicolumn{1}{c}{ [\ZnI]} &
 \multicolumn{1}{c}{ [\YII]} &
 \multicolumn{1}{c}{ [\ZrII]} &
 \multicolumn{1}{c}{ [\BaII]}   \\
 \multicolumn{1}{c|}{ HD} &
 \multicolumn{11}{|c}{ in $dex$}  \\
\hline 
\object{HD 4881}   &                     &                     &                     &                       & $-$0.64$\pm$0.10 (0/9)  &                     &                        &                     &                     &                     &                     \\
\object{HD 6028}   &                     & $-$0.48 (1)         &                     &                       & $+$0.02$\pm$0.19 (0/2)  &                     & $+$0.22 (0/1)          &                     &                     &                     &                     \\
\object{HD 17081}  &                     &                     &                     &                       & $-$0.21$\pm$0.08 (0/10) &                     & $-$0.10$\pm$0.02 (0/2) &                     &                     &                     &                     \\
\object{HD 17206}  & $+$0.03$\pm$0.03 (2)& $+$0.05 (1)         & $-$0.57 (1)         &                       & $-$0.01$\pm$0.16 (9/9)  & $-$0.04$\pm$0.08 (3)& $+$0.09$\pm$0.16 (17/0)& $-$0.35$\pm$0.15 (2)& $+$0.34$\pm$0.07 (2)& $+$0.77$\pm$0.02 (2)& $-$0.32$\pm$0.08 (3)\\
\object{HD 18256}  & $+$0.13$\pm$0.01 (2)& $-$0.12$\pm$0.05 (2)& $+$0.17 (1)         &                       & $+$0.22$\pm$0.12 (9/10) & $+$0.05 (1)         & $+$0.09$\pm$0.19 (23/0)&                     & $-$0.04$\pm$0.09 (3)&                     & $-$0.47$\pm$0.05 (2)\\
\object{HD 20010}  & $-$0.16$\pm$0.04 (4)& $-$0.47$\pm$0.13 (4)& $+$0.07$\pm$0.09 (6)&                       & $-$0.01$\pm$0.10 (8/11) & $+$0.12$\pm$0.21 (4)& $-$0.05$\pm$0.18 (26/0)& $-$0.14 (1)         & $-$0.02 (1)         &                     & $+$0.11$\pm$0.03 (2)\\
\object{HD 28978}  &                     & $-$0.52 (1)         & $+$0.18$\pm$0.15 (5)& $+$0.24$\pm$0.08 (0/3)& $+$0.06$\pm$0.08 (2/17) &                     & $+$0.30$\pm$0.14 (2/3) &                     & $+$0.49$\pm$0.07 (3)& $+$0.75$\pm$0.00 (3)& $+$0.67$\pm$0.03 (3)\\
\object{HD 31648}  &                     &                     &                     &                       &                         &                     &                        &                     & $+$0.75 (1)         &                     &                     \\
\object{HD 33564}  & $+$0.60 (1)         &                     & $-$0.21$\pm$0.10 (5)& $-$0.61 (1/0)         & $-$0.21$\pm$0.12 (5/9)  & $-$0.29$\pm$0.06 (6)& $-$0.13$\pm$0.11 (11/2)& $-$0.56$\pm$0.03 (2)& $+$0.05$\pm$0.11 (3)& $+$0.16$\pm$0.07 (5)& $+$0.69 (1)         \\
\object{HD 36112}  &                     &                     & $-$0.13 (1)         &                       & $-$0.11$\pm$0.10 (1/6)  & $-$0.16$\pm$0.15 (2)& $-$0.18 (1/0)          &                     & $+$0.55$\pm$0.11 (2)&                     &                     \\
\object{HD 95418}  &                     &                     &                     &                       & $+$0.49$\pm$0.06 (0/4)  &                     &                        &                     &                     &                     &                     \\
\object{HD 97633}  & $+$0.20$\pm$0.04 (2)&                     &                     &                       & $+$0.17$\pm$0.14 (0/7)  &                     &                        &                     &                     &                     &                     \\
\object{HD 100453} & $+$0.14$\pm$0.14 (3)&                     & $-$0.21$\pm$0.03 (3)& $+$0.28 (0/1)         & $+$0.03$\pm$0.10 (3/6)  & $-$0.04$\pm$0.17 (2)& $-$0.03$\pm$0.15 (7/0) & $-$0.52 (1)         &                     & $+$0.48$\pm$0.05 (2)& $-$0.02 (1)         \\
\object{HD 100546} &                     &                     &                     &                       & $-$0.87$\pm$0.05 (0/2)  &                     &                        &                     &                     &                     &                     \\
\object{HD 104237} & $+$0.07$\pm$0.10 (4)& $-$0.02$\pm$0.12 (4)& $+$0.08$\pm$0.05 (9)& $-$0.01$\pm$0.14 (1/2)& $+$0.17$\pm$0.13 (16/24)& $-$0.05$\pm$0.09 (7)& $+$0.09$\pm$0.14 (34/0)& $-$0.37$\pm$0.10 (2)& $+$0.18$\pm$0.10 (8)& $+$0.35$\pm$0.11 (6)& $+$0.31$\pm$0.05 (2)\\
\object{HD 135344} &                     &                     & $-$0.36 (1)         &                       & $+$0.06$\pm$0.06 (2/2)  &                     & $+$0.08$\pm$0.19 (2/0) & $-$0.45 (1)         &                     &                     &                     \\
\object{HD 139614} &                     &                     & $-$0.21$\pm$0.08 (5)& $-$0.01$\pm$0.08 (0/2)& $-$0.20$\pm$0.08 (11/19)& $-$0.30$\pm$0.01 (2)& $-$0.18$\pm$0.16 (13/0)& $-$0.52$\pm$0.14 (2)& $+$0.04$\pm$0.18 (3)& $+$0.47$\pm$0.04 (4)& $+$0.06$\pm$0.01 (2)\\
\object{HD 190073} &                     & $-$0.45 (1)         & $-$0.05$\pm$0.04 (3)& $+$0.09$\pm$0.08 (3/0)& $+$0.01$\pm$0.12 (2/17) &                     & $+$0.06$\pm$0.07 (0/2) &                     & $+$0.03 (1)         & $+$0.31 (1)         & $+$0.11 (1)         \\
\object{HD 250550} &                     &                     &                     &                       & $-$0.05 (0/1)           &                     & $+$0.06 (0/1)          &                     &                     &                     &                     \\
\hline                                                                                           
\object{Vega}         & $+$0.12&      & $-$0.77& $-$0.26& $-$0.48& $-$0.30& $+$0.02&      & $-$0.59& $-$0.93& $-$1.22\\
\object{$\lambda$ Boo}& $-$1.30&      & $-$0.70&        & $-$1.00&        & $-$1.20&      &        &        & $-$0.90\\
 ISM                  & $-$0.50&      &        &        &        &        &        &      &        &        &        \\
\hline	
\end{tabular}		           		
\end{center}		           		
\normalsize		                				       
\end{sidewaystable*}

Lines --in several cases only one-- of additional elements could be
detected in the spectra of some program stars. The results are as
follows. The He abundance of \object{HD 250550} was determined to be
[\HeI]=$-0.03$. For \object{HD 20010} we computed [\MnII]=$+0.11$ dex,
which is in good agreement with the derived \MnI \, abundance. For
\object{HD 104237} we determined [\AlII]=$+0.06$ dex and [\CuI]=$+0.26$ dex 
(for a solar abundance of 4.21 dex). For heavy elements, the following
results were obtained: the \LaII \, abundances for \object{HD 17206},
\object{HD 100453} and \object{HD 104237} are $+2.07$, $+1.46$ and
$+0.12$ dex, respectively; the Ce abundances for \object{HD 17206},
\object{HD 18256}, \object{HD 33564} and \object{HD 139614} amount to
$+1.26$, $+0.27$, $-0.15$ and $+0.64$ dex, respectively; for
\object{HD 17206}, we found [\NdII]=$-0.05$ dex, for \object{HD
  33564} [\NdII]=$+0.27$ dex, and for \object{HD 104237} 
[\NdII]=$+0.59$ dex; the abundance of samarium in \object{HD 104237}
is $+0.85$ dex; the europium abundance of \object{HD 17206} is $+0.23$
dex. The solar abundances of lanthanum, cerium, neodymium, samarium
and europium we used, are respectively $1.22$, $1.55$, $1.50$, $1.00$
and $0.51$ dex.

As a by-product of our spectral analysis, we have computed the projected
rotational velocity of each star in our sample. The values are listed
in Table~\ref{vsini}. The typical error is of the order of 10\%.

\subsection{Stars with evidence for (selective) depletion}

\subsubsection{\object{HD 4881}}

\citet{lagrangehenri90} considered this star, which has a Vega-like
IR excess, in a list of possible analogues to $\beta$ Pictoris in a
search for redshifted \CaII \, and \NaI \, absorption, which would point to
accretion of circumstellar matter; it is not clear from their paper
whether they observed the star. 
\citet{miroshnichenko} investigated the optical and IR photometry,
the H$\alpha$ emission and measured equivalent widths of prominent
spectral lines; they found that the star has a far IR excess, but
rejected it as a Herbig Ae/Be star and classified it as a classical Be
star. \citet{kalas02} discovered a Pleiades-like nebulosity around
\object{HD 4881}, but concluded that there was possibly still need for a second
IR source to explain the IRAS fluxes; they suggested that this extra
source consists of warm grains, either in a circumstellar disk or in a
broader region where nonequilibrium small-grain heating occurs.  

\begin{figure}
\rotatebox{0}{\resizebox{3.5in}{!}{\includegraphics{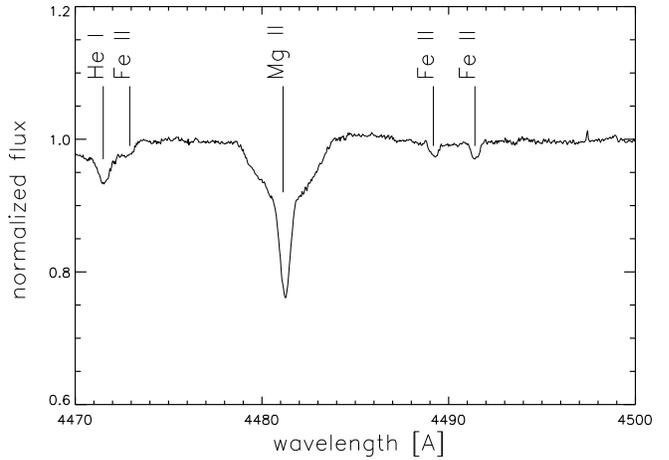}}}
\caption{ The peculiar shape of the \MgII \, line near 4481\AA\ in
  the spectrum of \object{HD 4881}. Four \SiII \, lines (4128, 4130, 6347, 6371\AA)
  and a \CaII \, line (3933\AA) show a similar profile.}
\label{magnesiumhr241}
\end{figure}

We observed \object{HD 4881} in the spectral region from 3860 to 5020\AA\ and from 5220 to 9100\AA.
A few strong spectral lines --\MgII \, $\lambda$4481, \SiII \,
$\lambda\lambda$4128, 4130, 6347, 6371 and \CaII \, $\lambda$3933-- in the
our spectrum have a peculiar shape (see
fig. \ref{magnesiumhr241}). The profile seems to be made up of two
components --- a superposition of a broad line and a narrower Gaussian
line. The straightforward interpretation is that the narrow line is of
circumstellar origin, while the broad line is the pressure broadened
photospheric line. A strange coincidence is then that the broadness of
the photospheric lines of the other elements in this deficient star is
close to that of the narrow components of the strong lines.
The projected rotational velocity we found for HD 4881 is 25
\kms. \citet{abt95} suggested a $v \sin i$ of 65 \kms, in rather
strong disagreement with our value. However, \citeauthor{abt95}
determined $v \sin i$ by fitting a Gaussian curve to the $\lambda$4481
\MgII \, line, of which the profile is more complex than Gaussian.

Our effective temperature for the Kurucz model of \object{HD 4881}, 11000K, is
consistent with the $B-V$ color index of $-$0.07 \citep{kalas02,flower96} and
spectral classification B9.5V. Higher effective temperatures 
lead to a discrepancy between the \FeI \, and the \FeII \, abundances. We thus
disagree with \citet{miroshnichenko} who proposed a spectral type
B8 for \object{HD 4881} and with the effective temperature of 12300K derived by
\citet{kalas02} from fitting the spectral energy
distribution. \citeauthor{miroshnichenko} also determined the effective
temperature on the basis of the Balmer jump and the H$\alpha$ line,
respectively $T_e$=10700$\pm$500K and 11000K; in agreement with our
determination.  

The result of our abundance analysis of \object{HD 4881} is that it is
deficient in all elements we studied, i.e. this star reveals a
complete depletion pattern. The iron abundance, measured from 4 \FeI \, 
lines and 19 \FeII \, lines, is about $-0.70$ dex. Similar deficiencies
are found for titanium and chromium, from 12 and 9 lines respectively,
and for the $\alpha$ element magnesium from 3 lines, while silicon (5
lines) appears still more deficient. 
With respect to the $\lambda$ Bootis phenomenon, however, we note that
also for carbon, nitrogen and oxygen subsolar abundances were found.  

An alternative explanation for the peculiar line shape
observed in a number of --intrinsically strong-- lines may be
binarity. When both components have comparable effective
temperatures, the spectrum of \object{HD 4881} is the
superposition of two spectra. Under this hypothesis, the line
profile displayed in Fig.~\ref{magnesiumhr241} contains a broad
feature belonging to the component of the binary with the largest
projected rotational velocity ($v \sin i \approx 100$ \kms). The
narrow absorption feature is then due to the second component, which also
causes the other narrow photospheric absorption lines in the composite
spectrum. The binary hypothesis potentially provides a solution for
the current inconsistencies in determinations of the effective
temperature and the projected rotational velocity of \object{HD
 4881}. Furthermore, the observed overall chemical depletion pattern
may be due to veiling effects. Although this hypothesis appears
very attractive, no independent observational evidence for the binary
nature of \object{HD 4881} is available in the literature.

\begin{table}
\caption{The projected rotational velocity. \label{vsini}}
\begin{center}
\begin{tabular}{cc|cc}
 \hline
 \multicolumn{4}{c}{\bf Projected rotational velocity}\\ \hline \hline
 \multicolumn{1}{c}{Star} &
 \multicolumn{1}{c|}{$v \sin i$} &
 \multicolumn{1}{|c}{Star} &
 \multicolumn{1}{c}{$v \sin i$}\\  
 \multicolumn{1}{c}{HD} &
 \multicolumn{1}{c|}{[\kms]} &
 \multicolumn{1}{|c}{HD} &
 \multicolumn{1}{c}{[\kms]}  \\
\hline 
\object{HD 4881}   &  25  & \object{HD 97048}  & 140 \\
\object{HD 6028}   &  95  & \object{HD 97633}  &  20 \\ 	
\object{HD 17081}  &  18  & \object{HD 100453} &  39 \\ 	
\object{HD 17206}  &  22  & \object{HD 100546} &  55 \\ 	
\object{HD 18256}  &  17  & \object{HD 102647} & 120 \\
\object{HD 20010}  &  10  & \object{HD 104237} &  10 \\
\object{HD 28978}  &  23  & \object{HD 135344} &  69 \\	
\object{HD 31293}  & 100  & \object{HD 139614} &  24 \\	
\object{HD 31648}  &  90  & \object{HD 190073} &   9 \\	
\object{HD 33564}  &  12  & \object{HD 244604} &  90 \\		  
\object{HD 36112}  &  60  & \object{HD 250550} &  54 \\	 
\object{HD 95418}  &  40  &                    &     \\
 \hline
\end{tabular}
\end{center}
\end{table}

\subsubsection{\object{AB Aur}}

\object{AB Aur} (\object{HD 31293}) has been previously mentioned as a $\lambda$
Bootis candidate by \citet{renson90}. Indeed, while we find slightly
suprasolar values for the nitrogen and oxygen abundances, iron appears
to be strongly depleted. Admittedly, the iron abundance is based on
only 2 lines, but these lines provided reliable abundances for the
other sample stars, and even taking into account the projected
rotational velocity of 100 \kms, it seems impossible to assign a solar
iron abundance to this frequently studied HAEBE star. On the other
hand, \object{AB Aur} appears to be only mildly deficient in silicon
([Si]=$-0.11$ dex). The possible $\lambda$ Bootis nature of this
object thus needs further investigation. 

\subsubsection{\object{HD 100546}}

\object{HD 100546} is one of the closest and brightest HAEBE stars. In IUE
observations, \citet{grady} detected redshifted absorption
features from iron resonance lines reminiscent of those observed for
\object{$\beta$ Pictoris} and concluded that also \object{HD 100546} undergoes a
bombardment by bodies of cometary size. The ISO
\citep[\textit{Infrared Space Observatory}][]{kessler} spectrum of \object{HD 100546}
is spectacular, in the sense that it presents prominent crystalline
silicate features with a spectrum closely resembling that of comet
Hale-Bopp \citep{malfait98b}.  
The stellar parameters that were used in our Kurucz model for \object{HD
100546} are taken from \citet{meeus01}. The projected rotational
velocity of \object{HD 100546} is 55 \kms, so that most lines in the optical
spectrum should be suitable for abundance analysis purposes. However,
it turns out that only a limited number of --intrinsically strong--
lines could be measured accurately, and that the results consistently
point to this star being deficient in several important elements. In
fact, \object{HD 100546} turns out to be the only clear $\lambda$ Bootis star
in our sample, being deficient about 1 order of magnitude in magnesium
(4 lines), silicon (6 lines) and iron (5 lines), but apparently having
solar CNO abundances. Late B stars showing relatively few suitable CNO
lines, the latter statement is essentially based on our determination
of the oxygen abundance, which was estimated based on the measured
equivalent widths of 10 spectral lines --amongst others six lines
from the multiplets around $\lambda$6156 and $\lambda$6454 -- which
led to a consistent result with a scatter of 0.10 dex. The nitrogen
abundance was computed using only the 7468\AA\ \NI \, line. The six
\SiII \, lines included the easily recognizable 6347\AA\ and 6371\AA\ lines;
the five iron lines are those at 4233, 4555, 4583, 5169 and 5275\AA,
the atomic parameters of which are quite reliable. A similar
deficiency was found for chromium, where the lines at $\lambda$4558
and $\lambda$4588 were used to compute [\CrII]=-0.87 dex. 

\subsubsection{\object{HD 139614}}

The star \object{HD 139614} is one of the objects in the sample for which the
largest amount of lines can be studied, thanks to a small projected
rotational velocity, which we found to be 24 \kms, in agreement with
\citet{dunkin97a}. According to \citet{meeus98}, it is likely that
the low $v \sin i$ is at least partly due to a near-pole-on
inclination of the star. The stellar parameters for \object{HD 139614} are
again extracted from \citet{meeus01}.

\begin{figure}
\rotatebox{0}{\resizebox{3.5in}{!}{\includegraphics{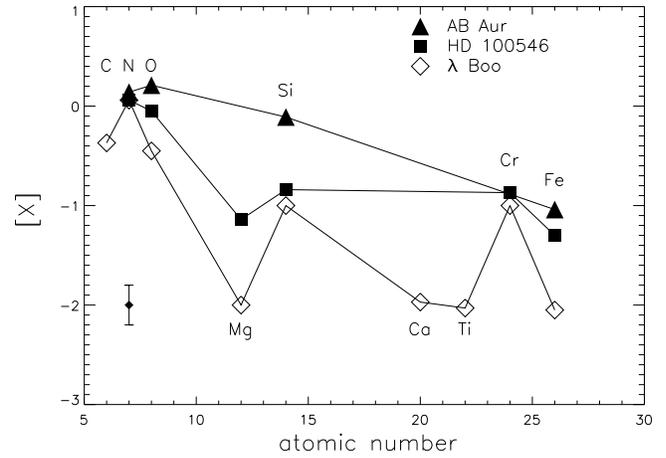}}}
\caption{ The elemental abundances of \object{AB Aur} and \object{HD 100546} determined
  in the present analysis compared to those of \object{$\lambda$ Boo}
  \citep{venn90}. The depletion pattern of \object{HD 100546} and \object{$\lambda$ Boo}
  are clearly similar. This is less obvious for \object{AB Aur}. In the lower
  left corner, the typical $\pm0.20$ dex error bar is shown.}
\label{lambdaBoopattern.ps}
\end{figure}

\begin{figure}
\rotatebox{0}{\resizebox{3.5in}{!}{\includegraphics{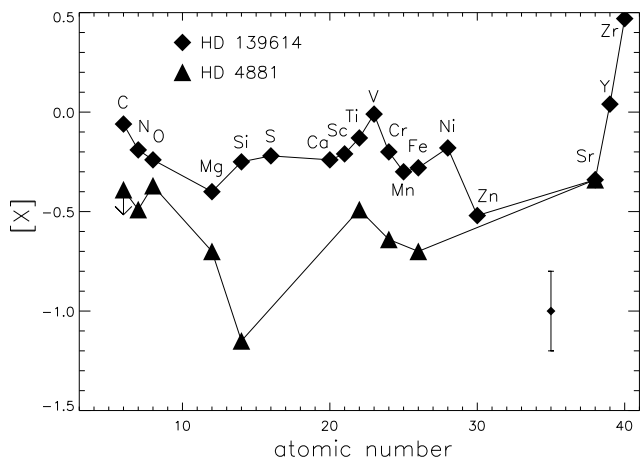}}}
\caption{ The elemental abundances of \object{HD 4881} and \object{HD
  139614} determined in the present analysis. Both patterns display an
  overall underabundance of $-0.65$ dex for \object{HD 4881} and
  $-0.22$ dex for \object{HD 139614}. In the lower right corner, the
  typical $\pm0.20$ dex error bar is shown.} 
\label{overalldepletion.ps}
\end{figure}

We found that all elements are slightly deficient
($\sim -$0.20 dex). The abundance pattern of this star is comparable to
the pattern in \object{HD 4881}, but the deficiency is less
strong. \citet{dunkin97a} also examined \object{HD 139614}
spectroscopically. They used a higher 
effective temperature $T_{\mathrm{e}}$=8250K and a lower gravity $\log g$=4.2
in their model. They concluded that [C]=$-0.07$ dex, [Mg]=$-0.08$ dex,
[Si]=$-0.52$ dex, [S]=$-0.12$ dex, [Ca]=$-0.33$ dex, [Ti]=$-0.21$ dex,
[Cr]=$-0.11$ dex, [Fe]=$-0.03$ dex and [Ni]=$-0.07$ dex. The
abundances of carbon, sulphur, calcium, titanium, chromium and nickel
are in relatively good agreement with our values. The computed
abundances of magnesium, silicon and iron differ from our results by
at least 0.15 dex. We note that \citeauthor{dunkin97a} base their
investigation mostly on strong lines, while we systematically
discarded all lines with an equivalent width exceeding 150 m\AA. In
fact, of the respectively 116 \FeI \, lines and 33 \FeII \, lines we used in
our analysis, only 1$+$4 occur in the list of lines studied by
\citet{dunkin97a}. It is not unlikely that our analysis is less
affected by the uncertainties induced by microturbulence, so that we
are confident that the small overall underabundance we found for
\object{HD 139614} is genuine. 
\\
\\
In Fig.~\ref{lambdaBoopattern.ps} the depletion pattern of \object{AB Aur} and
\object{HD 100546} is compared to the elemental-abundance pattern of \object{$\lambda$
Boo}. Fig.~\ref{overalldepletion.ps} displays the
overall-underabundance pattern of \object{HD 4481} and \object{HD
  139614}. No similarities with the elemental abundances of
\object{$\lambda$ Boo} are noted in these sources. 

\section{Discussion \label{discussion}}

A few possible mechanisms have been proposed to explain the $\lambda$ Bootis
abundance pattern. The selective-accretion theory of \citet{venn90} is
currently the most widely supported model. In the recent literature,
\citet{faraggiana99,gerbaldi03} and \citet{faraggiana04} mention that binarity
can play an underestimated role. The composite spectrum might mimic the spectrum
of a truly underabundant single star. Veiling due to the companion
will make the metallic lines in the spectrum appear weaker. It is
indeed not unlikely that some objects are misclassified as $\lambda$
Bootis stars, since the definition of this group of objects often used in the
literature is 'A/F stars with weak metallic lines'. We note however that the
$\lambda$ Boo abundance pattern also 
requires solar CNO and S abundances. Veiling due to binarity
works as much for the CNO and S absorption lines in the spectrum, hence
making the measured equivalent widths of these lines lower than
expected as well. The \textit{overall}
abundance pattern however remains roughly unaltered. It is
therefore important that metallic as well as CNO and 
S abundances are determined to classify an object as a
$\lambda$ Bootis star. We stress that the binarity of the selectively
depleted post-AGB stars cannot account for the observed
abundances either. This can be deduced from the solar abundance
CNO and S, and zinc in these objects \citep[e.g.][]{vanwinckel03}.

Even though young stars with circumstellar (debris) disks are a priori
fit environments for the selective accretion to take place
\citep[following][]{venn90},
the typical abundance pattern is not present in the vast majority of
these objects. Only one star in the present sample, 
\object{HD 100546}, appears to be selectively depleted. Because of the
similarities of the abundance pattern of \object{HD 100546} and
\object{$\lambda$ Boo}, we suggest that accretion of metal-depleted
gas is a plausible explanation for this source. Nevertheless, no other
HAEBE or Vega-type star in our sample displays this behaviour. A few possible
reasons can be proposed. Selective accretion might be a non-trivial
process, which only occurs in a minority of the sources. One of the
most stringent assumptions for the phenomenon to happen is the
decoupling of the gas and dust particles. When closely interacting,
the infalling, metal-depleted gas will drag the (metal-rich) dust
along. Both gas and dust will be accreted in this case, leaving the
photospheric abundances of the central star unaltered. A second
explanation for the lack of $\lambda$ Boo stars in this sample may be
the timescales involved. The dust particles are more easily
prevented from falling in, due to the larger radiation
pressure. Nevertheless, also the dust is expected to eventually spiral
in and be accreted. The depletion pattern is only present after the
gas has been accreted, and before the dust falls in onto the stellar
photosphere. The perceptibility of the $\lambda$ Bootis phenomenon
would then depend on the difference in accretion timescale between gas
and dust particles. Moreover, \citet{turcotte93} have modeled the
effect of stellar rotation and chemical diffusion on the photospheric
abundance pattern. They find that a constant infall of at least
$10^{-14} M_\odot\, \mathrm{yr}^{-1}$ is needed to maintain the depletion of the
metals when photospheric diffusion is taken into account. After accretion of gas
has stopped, the abundance pattern is destroyed on a timescale of
$10^{6}\, \mathrm{yr}$. Stellar rotation and the consequent
meridional circulation only plays a role for rotational velocities
above $\sim 125$ \kms. 

Imaging of \object{HD 100546} shows that the inclination of the
circumstellar disk around this source is $i \sim 50^\circ$
\citep{pantin00,augereau01,grady01} where $i=0^\circ$ represents
pole-on. Assuming that the 
rotational vector of the star and the disk are located along the same
axis, the rotational velocity of \object{HD 100546} is $v \approx 70$
\kms. This is consistent with \citet{turcotte93}, since the $\lambda$
Bootis abundance pattern is observable. \object{AB Aur} on the other
hand is a rapidly rotating star. Using the inclination of $i <
30^\circ$ \citep{eisner03,blake04} of the circumstellar disk for the stellar
rotation, the rotational velocity for this object is $v > 200$ \kms.
In this star, the photospheric nitrogen, oxygen and silicon are not
underabundant, indicating a normal abundance pattern. Strangely
enough, iron is clearly depleted, and this signature has not been
wiped out. Altogether, fast rotation alone cannot account for the
lack of $\lambda$ Boo stars in this sample; assuming an average
inclination angle of $i = 45^\circ$, only 5 sample sources (22\%) are
expected to have rotational velocities above 125 \kms.

HAEBE stars have ages of a few Myr and mass accretion is ongoing in
a large fraction of these sources. When the dust particles in the
circumstellar disk coagulate and grow, the coupling between gas and
dust becomes smaller. Therefore, one intuitively expects that the
$\lambda$ Bootis abundance pattern, which only occurs under stringent
conditions, appears in the latest phase of the pre-main-sequence
evolution of HAEBE stars (or later). During earlier stages, the mass
infall might be too vicious, dragging along gas and dust. A necessary,
but not satisfactory, condition for the depletion 
pattern to occur seems to be a quiet circumstellar environment, in
which the dust has grown to appreciable size or disappeared, and a
constant low-mass flow of depleted gas accretes onto the central star.

Vega-type stars are believed to be the successors of HAEBE stars. The
circumstellar matter around these stars consists of the
left-overs of the pre-main-sequence dust disk (cold
particles at large distances to the central star and possibly a
planetary system). None of our sample stars belonging to this group
displays selective depletion of the metals. Gas accretion in most of
these sources may have become too weak to produce a clear
underabundance pattern. 

\object{HD 100546} is probably at the end of its pre-main-sequence
lifetime. The observed $\lambda$ Bootis-like abundance pattern adds
another pecularity to the list of curiosa (e.g., high degree of
silicate crystallinity \citep{malfait98b},
spiral structure in the circumstellar disk \citep{grady01}) known
about this source. 

\section*{Appendix: notes on individual objects \label{appendix}}

\begin{figure}
\rotatebox{0}{\resizebox{3.5in}{!}{\includegraphics{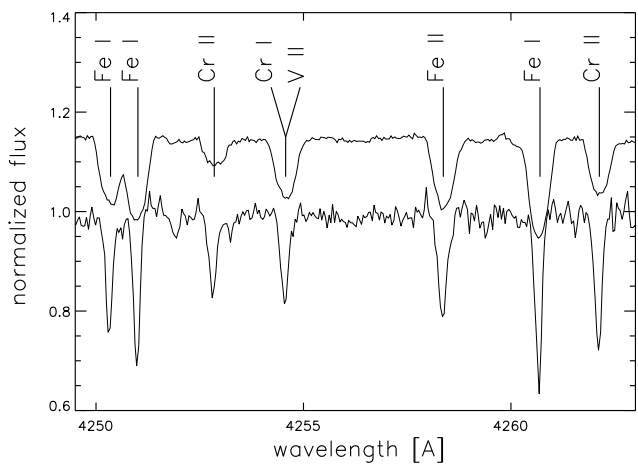}}}
\caption{ The spectra of HD 28978 (top), an A2 star with
  $T_{\mathrm{e}}$=9250K, and HD 190073 (bottom), are very similar. The
  agreement between both spectra confirms our lower --with respect to
  other authors-- effective
  temperature for HD 190073.} 
\label{hd190073hr1448}
\end{figure}

\textbf{HD 6028}. Stellar parameters come from \citet{coulson98}.\\
\textbf{HD 17081}. \citet{adelman91} determined the iron abundance
       [Fe]=$-0.07$ dex. \citet{takeda99} determined [O]=$-0.14$ dex,
       [Fe]=$-0.02$ dex and $v \sin i$=19.2 \kms. Our iron abundance
       is $\sim$0.20 dex lower. The oxygen abundance and the projected
       rotational velocity are in good agreement with our
       values. \citet{smith93} suggested [Cr]=$+0.33$ dex,
       [Fe]=$-0.02$ dex and [Ni]=$-0.05$ dex. Only the nickel
       abundance is in agreement with our determinations.  \\ 
\textbf{HD 17206}. \citet{marsakov95} derived an iron abundance of
$-0.07$ dex.  This is in excellent agreement with our results. \\ 
\textbf{HD 18256}. \citet{eggen98} computed [Fe]=$+ 0.13$ dex which is
fairly close to our determination. \\
\textbf{HD 20010}. \citet{santos01} and \citet{marsakov95} derived
respectively [Fe]=$-0.20$ dex and [Fe]=$-0.23$ dex . This strokes with our results. \\
\textbf{HD 28978}. The computed abundances of \citet{lemke89,lemke90}
       [\CI]=$-0.05$ dex, [\SiII]=$-0.06$ dex, [\CaI]=$-0.07$ dex,
       [\TiII]=$+0.15$ dex, [\FeI]=$+0.07$ dex, [\FeII]=$+0.19$ dex
       and [\BaII]=$+0.61$ dex are in fairly good agreement with our
       results. The result for the heavy element strontium differs
       from our determinations: [\SrII]$_{\mathrm{Lemke}}$=$+0.55$ dex
       versus our [\SrII]=$+0.31$ dex. \\ 
\textbf{HD 31648}. Stellar parameters are taken from
\citet{vandenancker98}. Our measured projected velocity agrees
reasonably with the value $v \sin i$=102 \kms of \citet{mora01}. \\  
\textbf{HD 33564}. \citet{deliyannis98} suggested [Fe]=$-0.05$
dex. \citet{marsakov95} derived [Fe]=$-0.11$ dex. We find a stronger
underabundance of iron, [\FeI]=$-0.35$ dex and [\FeII]=$-0.41$ dex. \\ 
\textbf{HD 36112}. \citet{beskrovnaya99} determined
       [Si]$\approx$$+0.40$ dex, [Ca]=$+0.06$ dex, [Fe]=$-0.02$ dex
       and $v \sin i$=60 \kms. The calcium and iron abundances and the
       projected rotational velocity are consistent with our
       findings. The silicon abundance of \citeauthor{beskrovnaya99}
       is $\sim$0.40 dex higher than our value.  \\ 
\textbf{HD 95418}. \citet{hill95} derived [Fe]=$+0.14$ dex which is
comparable to our results. \\ 
\textbf{HD 97048}. Stellar parameters come from
\citet{vandenancker98}. Because of the large projected rotational
velocity ($v \sin i$=140 \kms) the equivalent widths of the spectral
lines are difficult to determine. We based our oxygen abundance
computation on two \OI \, multiplets at 6156\AA\ and 6454\AA. \\
\textbf{HD 97633}. \citet{lemke89,lemke90} determined [\SiII]=$+0.15$ 
dex, [\CaI]=$+0.17$ dex, [\TiII]=$+0.20$ dex, [\FeI]=$+0.24$ dex and
[\FeII]=$+0.30$ dex. The silicon abundance is in good agreement with
our value. The other elemental abundances are systematically $\sim$
0.20 dex higher than our results. Our iron abundance ([Fe]=$+0.02$
dex) confirms the result of \citet{adelman88}. \citet{takeda99}
computed [O]=$-0.27$ dex, [Fe]=$+0.19$ dex, [Si]=$-0.43$ dex and $v
\sin i$=21.6 \kms. There is a discrepancy with our silicon and iron
abundances. \citet{smith93} determined [Cr]=$+0.43$ dex and
[Fe]=$+0.03$ dex. We find a chromium abundance of $+0.17$ dex. \\ 
\textbf{HD 100453}. Stellar parameters were taken from
\citet{meeus01}. We altered the $\log g$ value from 4.5 to 4.0. About
180 spectral lines were used in the abundance determinations.\\ 
\textbf{HD 102647}. Stellar parameters were extracted from
\citet{habing00}. \\ 
\textbf{HD 104237}. We took the stellar parameters of
\citet{meeus01}. Close to 440 lines were included in the abundance
computations.\\ 
\textbf{HD 135344}. We adopted the gravity parameter of
\citet{meeus01} from $\log g$=4.5 down to 4.0. We kept
$T_{\mathrm{e}}$ from that article. The projected rotational velocity
is 80 \kms. We specifically looked for unblended spectral lines by
comparing the spectrum of HD 135344 with the spectra of HD 18256 and
HD 20010, two stars with a similar effective temperature, but much
lower $v \sin i$. By doing this we diminish the risk of misidentifying
lines and measuring wrong equivalent widths. \\ 
\textbf{HD 190073}. Based on the A3IIIe spectral classification by
\citet{gray} and the A2IVe classification by \citet{mora01} we chose
an effective temperature $T_{\mathrm{e}}$ = 9250 K and a gravity $\log
g$ = 3.5 for the Kurucz model. Other articles suggested a much higher
effective temperature (e.g. 11120K and a B9IV classification,
\citet{cidale01} or A0IV by \citet{the}). Nevertheless, the lower
temperature was consistent with the spectrum of HD 190073 (see
fig. \ref{hd190073hr1448}).  
Some of the spectral lines of \CaI, \CaII, \ScII, \TiII, \CrII, \FeI \, 
and \FeII \, depict emission features. The emission fills the core of the
photospheric absorption line. Hence, the equivalent width of the line
cannot be measured accurately. The spectral lines that display
emission are therefore not included in the abundance analysis of HD
190073. \\ 
\textbf{HD 244604}. The stellar parameters of HD 244604 are extracted
from \citet{miroshnichenko}. \\ 
\textbf{HD 245185}. Not one single absorption line was detected. Our
signal-to-noise level in this spectrum was about S/N$\sim$30, which
makes it by far the worst spectrum in our sample. The projected
rotational velocity could not be derived. Stellar parameters for HD
245185 come from \citet{testi98}. \\ 
\textbf{HD 250550}. Stellar parameters were taken from
\citet{testi98}. \\ 

\bibliographystyle{aa}
\bibliography{/STER/55/bram/REFERENCES/references.bib}

\begin{thebibliography}{76}
\expandafter\ifx\csname natexlab\endcsname\relax\def\natexlab#1{#1}\fi

\bibitem[{{Abt} \& {Morrell}(1995)}]{abt95}
{Abt}, H.~A. \& {Morrell}, N.~I. 1995, \apjs, 99, 135

\bibitem[{{Adelman}(1988)}]{adelman88}
{Adelman}, S.~J. 1988, \mnras, 230, 671

\bibitem[{{Adelman} {et~al.}(1991){Adelman}, {Bolcal}, {Kocer}, \&
  {Hill}}]{adelman91}
{Adelman}, S.~J., {Bolcal}, C., {Kocer}, D., \& {Hill}, G. 1991, \mnras, 252,
  329

\bibitem[{{Anders} \& {Grevesse}(1989)}]{anders89}
{Anders}, E. \& {Grevesse}, N. 1989, \gca, 53, 197

\bibitem[{{Augereau} {et~al.}(2001){Augereau}, {Lagrange}, {Mouillet}, \&
  {M{\'e}nard}}]{augereau01}
{Augereau}, J.~C., {Lagrange}, A.~M., {Mouillet}, D., \& {M{\'e}nard}, F. 2001,
  \aap, 365, 78

\bibitem[{{Backman} \& {Paresce}(1993)}]{backman93}
{Backman}, D.~E. \& {Paresce}, F. 1993, in Protostars and Planets III,
  1253--1304

\bibitem[{{Bell} \& {Lin}(1994)}]{bell94}
{Bell}, K.~R. \& {Lin}, D.~N.~C. 1994, \apj, 427, 987

\bibitem[{{Beskrovnaya} {et~al.}(1999){Beskrovnaya}, {Pogodin},
  {Miroshnichenko}, {Th{\'e}}, {Savanov}, {Shakhovskoy}, {Rostopchina},
  {Kozlova}, \& {Kuratov}}]{beskrovnaya99}
{Beskrovnaya}, N.~G., {Pogodin}, M.~A., {Miroshnichenko}, A.~S., {et~al.} 1999,
  \aap, 343, 163

\bibitem[{{Bhatt} \& {Manoj}(2000)}]{bhatt00}
{Bhatt}, H.~C. \& {Manoj}, P. 2000, \aap, 362, 978

\bibitem[{{Blake} \& {Boogert}(2004)}]{blake04}
{Blake}, G.~A. \& {Boogert}, A.~C.~A. 2004, \apjl, 606, L73

\bibitem[{{Chiang} \& {Goldreich}(1997)}]{chiang}
{Chiang}, E.~I. \& {Goldreich}, P. 1997, \apj, 490, 368

\bibitem[{{Cidale} {et~al.}(2001){Cidale}, {Zorec}, \&
  {Tringaniello}}]{cidale01}
{Cidale}, L., {Zorec}, J., \& {Tringaniello}, L. 2001, \aap, 368, 160

\bibitem[{{Coulson} {et~al.}(1998){Coulson}, {Walther}, \& {Dent}}]{coulson98}
{Coulson}, I.~M., {Walther}, D.~M., \& {Dent}, W.~R.~F. 1998, \mnras, 296, 934

\bibitem[{{Deliyannis} {et~al.}(1998){Deliyannis}, {Boesgaard}, {Stephens},
  {King}, {Vogt}, \& {Keane}}]{deliyannis98}
{Deliyannis}, C.~P., {Boesgaard}, A.~M., {Stephens}, A., {et~al.} 1998, \apjl,
  498, L147+

\bibitem[{{Dullemond}(2002)}]{dullemond02}
{Dullemond}, C.~P. 2002, \aap, 395, 853

\bibitem[{{Dullemond} \& {Dominik}(2004)}]{dullemond04}
{Dullemond}, C.~P. \& {Dominik}, C. 2004, \aap, 417, 159

\bibitem[{{Dullemond} {et~al.}(2001){Dullemond}, {Dominik}, \&
  {Natta}}]{dullemond01}
{Dullemond}, C.~P., {Dominik}, C., \& {Natta}, A. 2001, \apj, 560, 957

\bibitem[{{Dunkin} {et~al.}(1997){Dunkin}, {Barlow}, \& {Ryan}}]{dunkin97a}
{Dunkin}, S.~K., {Barlow}, M.~J., \& {Ryan}, S.~G. 1997, \mnras, 286, 604

\bibitem[{{Eggen}(1998)}]{eggen98}
{Eggen}, O.~J. 1998, \aj, 115, 2397

\bibitem[{{Eisner} {et~al.}(2003){Eisner}, {Lane}, {Akeson}, {Hillenbrand}, \&
  {Sargent}}]{eisner03}
{Eisner}, J.~A., {Lane}, B.~F., {Akeson}, R.~L., {Hillenbrand}, L.~A., \&
  {Sargent}, A.~I. 2003, \apj, 588, 360

\bibitem[{{Erspamer} \& {North}(2002)}]{erspamer02}
{Erspamer}, D. \& {North}, P. 2002, \aap, 383, 227

\bibitem[{{Faraggiana} \& {Bonifacio}(1999)}]{faraggiana99}
{Faraggiana}, R. \& {Bonifacio}, P. 1999, \aap, 349, 521

\bibitem[{{Faraggiana} {et~al.}(2004){Faraggiana}, {Bonifacio}, {Caffau},
  {Gerbaldi}, \& {Nonino}}]{faraggiana04}
{Faraggiana}, R., {Bonifacio}, P., {Caffau}, E., {Gerbaldi}, M., \& {Nonino},
  M. 2004, \aap accepted, astro-ph/0406265

\bibitem[{{Finkenzeller} \& {Mundt}(1984)}]{finkenzeller84}
{Finkenzeller}, U. \& {Mundt}, R. 1984, \aaps, 55, 109

\bibitem[{{Flower}(1996)}]{flower96}
{Flower}, P.~J. 1996, \apj, 469, 355

\bibitem[{{Fuente} {et~al.}(2003){Fuente}, {Rodr{\'{\i}}guez-Franco}, {Testi},
  {Natta}, {Bachiller}, \& {Neri}}]{fuente}
{Fuente}, A., {Rodr{\'{\i}}guez-Franco}, A., {Testi}, L., {et~al.} 2003, \apjl,
  598, L39

\bibitem[{{Gerbaldi} {et~al.}(2003){Gerbaldi}, {Faraggiana}, \&
  {Lai}}]{gerbaldi03}
{Gerbaldi}, M., {Faraggiana}, R., \& {Lai}, O. 2003, \aap, 412, 447

\bibitem[{{Grady} {et~al.}(1995){Grady}, {P{\'e}rez}, {Th{\'e}}, {Grinin}, {de
  Winter}, {Johnson}, \& {Talavera}}]{grady}
{Grady}, C.~A., {P{\'e}rez}, M.~R., {Th{\'e}}, P.~S., {et~al.} 1995, \aap, 302,
  472

\bibitem[{{Grady} {et~al.}(2001){Grady}, {Polomski}, {Henning}, {Stecklum},
  {Woodgate}, {Telesco}, {Pi{\~n}a}, {Gull}, {Boggess}, {Bowers}, {Bruhweiler},
  {Clampin}, {Danks}, {Green}, {Heap}, {Hutchings}, {Jenkins}, {Joseph},
  {Kaiser}, {Kimble}, {Kraemer}, {Lindler}, {Linsky}, {Maran}, {Moos}, {Plait},
  {Roesler}, {Timothy}, \& {Weistrop}}]{grady01}
{Grady}, C.~A., {Polomski}, E.~F., {Henning}, T., {et~al.} 2001, \aj, 122, 3396

\bibitem[{{Gray} \& {Corbally}(1998)}]{gray}
{Gray}, R.~O. \& {Corbally}, C.~J. 1998, \aj, 116, 2530

\bibitem[{{Habing} {et~al.}(2000){Habing}, {Dominik}, {Jourdain de Muizon},
  {Laureijs}, {Kessler}, {Leech}, {Metcalfe}, {Siebenmorgen}, {Trams}, \&
  {Bouchet}}]{habing00}
{Habing}, H.~J., {Dominik}, C., {Jourdain de Muizon}, M., {et~al.} 2000, VizieR
  Online Data Catalog, 336, 50545

\bibitem[{{Heiter}(2002)}]{heiter02}
{Heiter}, U. 2002, \aap, 381, 959

\bibitem[{{Herbig}(1960)}]{herbig60}
{Herbig}, G.~H. 1960, \apjs, 4, 337

\bibitem[{{Hill}(1995)}]{hill95}
{Hill}, G.~M. 1995, \aap, 294, 536

\bibitem[{{Jenkins}(1989)}]{jenkins89}
{Jenkins}, E. 1989, in IAU Symp. 135: Interstellar Dust, 23

\bibitem[{{Kalas} {et~al.}(2002){Kalas}, {Graham}, {Beckwith}, {Jewitt}, \&
  {Lloyd}}]{kalas02}
{Kalas}, P., {Graham}, J.~R., {Beckwith}, S.~V.~W., {Jewitt}, D.~C., \&
  {Lloyd}, J.~P. 2002, \apj, 567, 999

\bibitem[{{Kenyon} \& {Hartmann}(1987)}]{kenyon87}
{Kenyon}, S.~J. \& {Hartmann}, L. 1987, \apj, 323, 714

\bibitem[{{Kessler} {et~al.}(1996){Kessler}, {Steinz}, {Anderegg}, {Clavel},
  {Drechsel}, {Estaria}, {Faelker}, {Riedinger}, {Robson}, {Taylor}, \&
  {Ximenez de Ferran}}]{kessler}
{Kessler}, M.~F., {Steinz}, J.~A., {Anderegg}, M.~E., {et~al.} 1996, \aap, 315,
  L27

\bibitem[{{Lagrange-Henri} {et~al.}(1990){Lagrange-Henri}, {Ferlet},
  {Vidal-Madjar}, {Beust}, {Gry}, \& {Lallement}}]{lagrangehenri90}
{Lagrange-Henri}, A.~M., {Ferlet}, R., {Vidal-Madjar}, A., {et~al.} 1990,
  \aaps, 85, 1089

\bibitem[{{Lambert} {et~al.}(1988){Lambert}, {Hinkle}, \& {Luck}}]{lambert88}
{Lambert}, D.~L., {Hinkle}, K.~H., \& {Luck}, R.~E. 1988, \apj, 333, 917

\bibitem[{{Lemke}(1989)}]{lemke89}
{Lemke}, M. 1989, \aap, 225, 125

\bibitem[{{Lemke}(1990)}]{lemke90}
---. 1990, \aap, 240, 331

\bibitem[{{Lin} \& {Papaloizou}(1980)}]{lin80}
{Lin}, D.~N.~C. \& {Papaloizou}, J. 1980, \mnras, 191, 37

\bibitem[{{Malfait} {et~al.}(1998{\natexlab{a}}){Malfait}, {Bogaert}, \&
  {Waelkens}}]{malfait}
{Malfait}, K., {Bogaert}, E., \& {Waelkens}, C. 1998{\natexlab{a}}, \aap, 331,
  211

\bibitem[{{Malfait} {et~al.}(1998{\natexlab{b}}){Malfait}, {Waelkens},
  {Waters}, {Vandenbussche}, {Huygen}, \& {de Graauw}}]{malfait98b}
{Malfait}, K., {Waelkens}, C., {Waters}, L.~B.~F.~M., {et~al.}
  1998{\natexlab{b}}, \aap, 332, L25

\bibitem[{{Mannings} \& {Sargent}(1997)}]{mannings97}
{Mannings}, V. \& {Sargent}, A.~I. 1997, \apj, 490, 792

\bibitem[{{Marsakov} \& {Shevelev}(1995)}]{marsakov95}
{Marsakov}, V.~A. \& {Shevelev}, Y.~G. 1995, Bulletin d'Information du Centre
  de Donnees Stellaires, 47, 13

\bibitem[{{Meeus} {et~al.}(1998){Meeus}, {Waelkens}, \& {Malfait}}]{meeus98}
{Meeus}, G., {Waelkens}, C., \& {Malfait}, K. 1998, \aap, 329, 131

\bibitem[{{Meeus} {et~al.}(2001){Meeus}, {Waters}, {Bouwman}, {van den Ancker},
  {Waelkens}, \& {Malfait}}]{meeus01}
{Meeus}, G., {Waters}, L.~B.~F.~M., {Bouwman}, J., {et~al.} 2001, \aap, 365,
  476

\bibitem[{{Miroshnichenko} {et~al.}(1999){Miroshnichenko}, {Mulliss},
  {Bjorkman}, {Morrison}, {Kuratov}, \& {Wisniewski}}]{miroshnichenko}
{Miroshnichenko}, A.~S., {Mulliss}, C.~L., {Bjorkman}, K.~S., {et~al.} 1999,
  \mnras, 302, 612

\bibitem[{{Molster} {et~al.}(1999){Molster}, {Waters}, {Trams}, {Van Winckel},
  {Decin}, {van Loon}, {J{\" a}ger}, {Henning}, {K{\" a}ufl}, {de Koter}, \&
  {Bouwman}}]{molster99}
{Molster}, F.~J., {Waters}, L.~B.~F.~M., {Trams}, N.~R., {et~al.} 1999, \aap,
  350, 163

\bibitem[{{Mora} {et~al.}(2001){Mora}, {Mer{\'{\i}}n}, {Solano}, {Montesinos},
  {de Winter}, {Eiroa}, {Ferlet}, {Grady}, {Davies}, {Miranda}, {Oudmaijer},
  {Palacios}, {Quirrenbach}, {Harris}, {Rauer}, {Cameron}, {Deeg},
  {Garz{\'o}n}, {Penny}, {Schneider}, {Tsapras}, \& {Wesselius}}]{mora01}
{Mora}, A., {Mer{\'{\i}}n}, B., {Solano}, E., {et~al.} 2001, \aap, 378, 116

\bibitem[{{Natta} {et~al.}(2004){Natta}, {Testi}, {Neri}, {Shepherd}, \&
  {Wilner}}]{natta04}
{Natta}, A., {Testi}, L., {Neri}, R., {Shepherd}, D.~S., \& {Wilner}, D.~J.
  2004, \aap, 416, 179

\bibitem[{{Pantin} {et~al.}(2000){Pantin}, {Waelkens}, \& {Lagage}}]{pantin00}
{Pantin}, E., {Waelkens}, C., \& {Lagage}, P.~O. 2000, \aap, 361, L9

\bibitem[{{Patten} \& {Willson}(1991)}]{patten91}
{Patten}, B.~M. \& {Willson}, L.~A. 1991, \aj, 102, 323

\bibitem[{{Paunzen} {et~al.}(1999){Paunzen}, {Andrievsky}, {Chernyshova},
  {Klochkova}, {Panchuk}, \& {Handler}}]{paunzen99}
{Paunzen}, E., {Andrievsky}, S.~M., {Chernyshova}, I.~V., {et~al.} 1999, \aap,
  351, 981

\bibitem[{{P{\'e}rez} \& {Grady}(1997)}]{perez}
{P{\'e}rez}, M.~R. \& {Grady}, C.~A. 1997, Space Science Reviews, 82, 407

\bibitem[{{Pi{\'e}tu} {et~al.}(2003){Pi{\'e}tu}, {Dutrey}, \& {Kahane}}]{pietu}
{Pi{\'e}tu}, V., {Dutrey}, A., \& {Kahane}, C. 2003, \aap, 398, 565

\bibitem[{{Qiu} {et~al.}(2001){Qiu}, {Zhao}, {Chen}, \& {Li}}]{qiu01}
{Qiu}, H.~M., {Zhao}, G., {Chen}, Y.~Q., \& {Li}, Z.~W. 2001, \apj, 548, 953

\bibitem[{{Renson} {et~al.}(1990){Renson}, {Faraggiana}, \& {Boehm}}]{renson90}
{Renson}, P., {Faraggiana}, R., \& {Boehm}, C. 1990, Bulletin d'Information du
  Centre de Donnees Stellaires, 38, 137

\bibitem[{{Santos} {et~al.}(2001){Santos}, {Israelian}, \& {Mayor}}]{santos01}
{Santos}, N.~C., {Israelian}, G., \& {Mayor}, M. 2001, \aap, 373, 1019

\bibitem[{{Smith} \& {Dworetsky}(1993)}]{smith93}
{Smith}, K.~C. \& {Dworetsky}, M.~M. 1993, \aap, 274, 335

\bibitem[{{Sneden}(1973)}]{sneden73}
{Sneden}, C.~A. 1973, Ph.D.~Thesis

\bibitem[{{Song} {et~al.}(2000){Song}, {Caillault}, {Barrado y Navascu{\'e}s},
  {Stauffer}, \& {Randich}}]{song00}
{Song}, I., {Caillault}, J.-P., {Barrado y Navascu{\'e}s}, D., {Stauffer},
  J.~R., \& {Randich}, S. 2000, \apjl, 533, L41

\bibitem[{{Takeda} {et~al.}(1999){Takeda}, {Takada-Hidai}, {Jugaku}, {Sakaue},
  \& {Sadakane}}]{takeda99}
{Takeda}, Y., {Takada-Hidai}, M., {Jugaku}, J., {Sakaue}, A., \& {Sadakane}, K.
  1999, \pasj, 51, 961

\bibitem[{{Testi} {et~al.}(2003){Testi}, {Natta}, {Shepherd}, \&
  {Wilner}}]{testi}
{Testi}, L., {Natta}, A., {Shepherd}, D.~S., \& {Wilner}, D.~J. 2003, \aap,
  403, 323

\bibitem[{{Testi} {et~al.}(1998){Testi}, {Palla}, \& {Natta}}]{testi98}
{Testi}, L., {Palla}, F., \& {Natta}, A. 1998, \aaps, 133, 81

\bibitem[{{Th{\'e}} {et~al.}(1994){Th{\'e}}, {de Winter}, \& {P{\'e}rez}}]{the}
{Th{\'e}}, P.~S., {de Winter}, D., \& {P{\'e}rez}, M.~R. 1994, \aaps, 104, 315

\bibitem[{{Turcotte} \& {Charbonneau}(1993)}]{turcotte93}
{Turcotte}, S. \& {Charbonneau}, P. 1993, \apj, 413, 376

\bibitem[{{van den Ancker} {et~al.}(1998){van den Ancker}, {de Winter}, \&
  {Tjin A Djie}}]{vandenancker98}
{van den Ancker}, M.~E., {de Winter}, D., \& {Tjin A Djie}, H.~R.~E. 1998,
  \aap, 330, 145

\bibitem[{{van Winckel}(2003)}]{vanwinckel03}
{van Winckel}, H. 2003, \araa, 41, 391

\bibitem[{{Van Winckel} {et~al.}(1995){Van Winckel}, {Waelkens}, \&
  {Waters}}]{vanwinckel95}
{Van Winckel}, H., {Waelkens}, C., \& {Waters}, L.~B.~F.~M. 1995, \aap, 293,
  L25

\bibitem[{{Venn} \& {Lambert}(1990)}]{venn90}
{Venn}, K.~A. \& {Lambert}, D.~L. 1990, \apj, 363, 234

\bibitem[{{Waters} {et~al.}(1998){Waters}, {Beintema}, {Zijlstra}, {de Koter},
  {Molster}, {Bouwman}, {de Jong}, {Pottasch}, \& {de Graauw}}]{waters98b}
{Waters}, L.~B.~F.~M., {Beintema}, D.~A., {Zijlstra}, A.~A., {et~al.} 1998,
  \aap, 331, L61

\bibitem[{{Waters} {et~al.}(1992){Waters}, {Trams}, \& {Waelkens}}]{waters92}
{Waters}, L.~B.~F.~M., {Trams}, N.~R., \& {Waelkens}, C. 1992, \aap, 262, L37

\bibitem[{{Waters} \& {Waelkens}(1998)}]{waters98}
{Waters}, L.~B.~F.~M. \& {Waelkens}, C. 1998, \araa, 36, 233

\end{thebibliography}

\end{document}